\documentclass[review,1p,doublespacing]{elsarticle}
\usepackage{graphics}
\usepackage[ruled]{algorithm2e}
\usepackage{graphicx}
\usepackage{epsfig}

\usepackage{amssymb,amsmath,amsfonts}
\usepackage{amsthm}

\begin{document}

\begin{frontmatter}



\title{A simple mathematical model for assessment of anti-toxin antibodies}


\author{Alex Skvortsov\footnote{Email:~Alex.Skvortsov@dsto.defence.gov.au, Fax: + 61 3 96268410}}\address{HPP Division, Defence Science and Technology Organisation,
506 Lorimer Street, Fishermans Bend, Vic 3207, Australia }
\author{Peter Gray\footnote{Email:~Peter.Gray@dsto.defence.gov.au}}\address{HPP Division, Defence Science and Technology Organisation, 506 Lorimer Street, Fishermans Bend, Vic 3207, Australia}

%
%

\begin{abstract}
The toxins associated with infectious diseases are potential targets
for inhibitors which have the potential for prophylactic or
therapeutic use. Many antibodies have been generated for this
purpose, and the objective of this study was to develop a simple mathematical model that may be used to evaluate the potential
protective effect of antibodies. This model was used to evaluate the
contributions of antibody affinity and concentration to reducing
antibody-receptor complex formation and internalization. The model
also enables prediction of the antibody kinetic constants and
concentration required to provide a specified degree of protection.
We hope that this  model, once validated experimentally, will be a useful tool for in vitro selection of potentially protective antibodies for
progression to in vivo evaluation.

\end{abstract}


\end{frontmatter}



\section{Introduction}

\label{} Passive immunization using antibodies has been used
successfully for treatment and prophylaxis of infectious disease in
humans  \cite{R1}  and there is increasing interest in the use of
antibodies for treatment of infectious diseases  that may be used as
terrorist weapons, but for which the risk is not sufficiently high
to justify preventive vaccination of a large civilian population
\cite{R2}. Toxins are an important potential target for designing
therapies against these threats and a broad range of approaches has
been taken to develop inhibitors that may be of prophylactic or
therapeutic use \cite{R3}.

Antibody engineering techniques allow affinity maturation of
antibodies and these techniques are being exploited to produce
inhibitors for a number of toxins \cite{R4}, \cite{R5}. The emphasis of this
approach is on producing reagents with high affinity, based on the
proposition that higher affinity will provide better protection.

However affinity, by itself, is a poor predictor of protective or
therapeutic potential. Antibodies with high in vitro affinity for
toxins do not automatically confer protection in vivo \cite{R6}, \cite{R7} and may exacerbate the toxicity \cite{R8}, \cite{R9}.  The effects of using multiple antibodies with high affinities may be additive \cite{R10} or synergistic \cite{R6} or without effect \cite{R7}. In addition, epitope specificity \cite{R11},
antibody titre \cite{R12} - \cite{R16}  and dissociation rate \cite{R17} have been correlated with protection.

Toxins are produced by a number of plants, animals and
microorganisms. Toxins may act at the cell surface and either damage
the cytoplasmic membrane or bind to a receptor and act via
transmembrane signalling subsequent to that binding \cite{R18}.
Alternatively, toxins may cross the cell membrane and act on
intracellular targets \cite{R18}. For example: anthrax lethal toxin, ricin
and cholera toxin bind to a cell surface receptor and make use of
cellular membrane trafficking to enter the cell \cite{R19}, \cite{R20}.

The objective of this study is to develop a simple mathematical
model that may be used to predict the optimum antibody parameters
(kinetic constants and concentration) needed to inhibit the binding
of the toxin to its receptor. These predictions may be used to
select candidate antibodies for progression to in vivo evaluation
and to assess the potential value of affinity enhancement.

This paper is an extension to our previous work \cite{AR21}. In the model presented below we explicitly take into account the process of toxin internalization and diffusive fluxes around the cell.

\begin{figure}
    \centering
    \includegraphics[height= 8cm,width=0.8\textwidth]{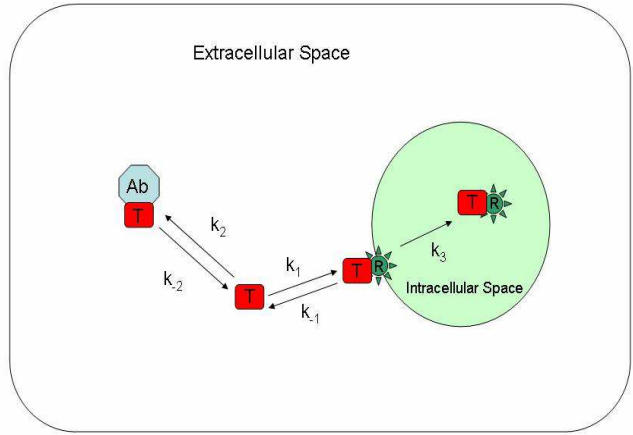}
\caption{Schematic representation of the model for receptor-toxin-antibody interaction.}
\label{F:0}
\end{figure}

\section{Model}

The kinetic model describing the interactions of toxins with cell
receptors can be formulated based on the well-known analytical framework
for ligand-receptor binding. The models of this process have been
studied for many years and  a vast amount of literature
has accumulated on this subject (see \cite{A1}-- \cite{A5}  and refs therein).

When a toxin diffuses in the extracellular environment and binds to the cell surface receptors,
the toxin concentration will vary both in space and time. Any rigorous description of this process would entail a system of Partial Differential Equations (PDE), which couples extracellular
diffusion with reaction kinetics of the cell surface. The resulting system of PDE is nonlinear and too complex to be treated analytically. This complexity makes unfeasible any comprehensive study of parameter optimization. From another perspective, it is well known that under some rather broad conditions (see \cite{A1} - \cite{A5} and refs)  the reaction-diffusion system of  the ligand-receptor binding can be well approximated by a system of Ordinary Differential Equations in which the spatial variability of the process is simulated by different concentrations of species in initially predefined spatial domains  (called compartments). Although, this compartment model is significantly simpler than the initial reaction-diffusion system, it still allows a consistent description of diffusion fluxes \cite{A2}, \cite{A4}, \cite{A5}. In the current paper we  use the compartment-model approach for our analytical study and numerical simulations.

To begin, we consider the following simple model. The toxin, $T$, binds
reversibly to cell surface receptors, $R$, with a forward rate $k_1$
and a reverse rate $k_{-1}$ to form the toxin-receptor complex $C_R$
which is then slowly internalized at a rate $k_3$ . The neutralizing
antibody binds competitively to the toxin with on and off rates of
$k_2$ and $k_{-2}$ respectively. The antibody-toxin complex, $C_A$,
remains in the extracellular space (see Fig.\ref{F:0}).

We can easily write an equation for the toxin-receptor binding (viz, without antibody being present). For a spherical cell of radius $a$ with the toxin binding to its surface \cite{A1}- \cite{A5}
\begin{eqnarray}
\label{m:eq1} \frac{d C_R}{dt} = {k^{e}_f} R T   + k^{e}_{r} C_R ,
\end{eqnarray}
where $C_R$ is the concentration  of the bound receptors
(toxin-receptor complexes), $R$ is the
concentration of receptors,  $T$ is the bulk toxin concentration
(i.e. far from the cell surface) and is assumed to be spatially
uniform. The effective forward and reverse rate coefficients are
defined by \cite{A1} - \cite{A5}
\begin{eqnarray}
\label{m:eq2}  k^{e}_{f} =  \gamma k_{1} , ~  k^{e}_{r} = \gamma k_{-1},
\end{eqnarray}
where  $k_{1}, k_{-1}$ are
intrinsic reaction rates, $k_D = 4 \pi a D$ is the diffusion rate,
$D$ is the diffusivity of toxin in the extracellular space, $\gamma = 1/( 1 + R k_{1}/k_D )\leq 1$ \cite{A2},\cite{A3},\cite{A5}.

The bulk concentration of toxin $T$ is mainly driven by the binding to
antibody. Therefore, in this case we can write an equation system similar to (\ref{m:eq1}), but
without any ``diffusive'' modification of the intrinsic rate constants:
\begin{eqnarray}
\label{m:eq3} \frac{d C_A}{dt} = {k_2} A T   + k_{-2} C_A ,
\end{eqnarray}
where $C_A$ is the concentration of toxin-antibody complexes, $A$ is the concentration of antibody.

The process of toxin internalization is phenomenologically introduced into our model by
the equation
\begin{eqnarray}
\label{m:eq31}
 \frac{d T_i}{dt} = {k_3} C_R ,
\end{eqnarray}
where $T_i$ is the concentration of internalized toxin. The
corresponding term should be included in (\ref{m:eq1}), so we arrive
at  modified expression for $k^{e}_{r}$
\begin{eqnarray}
\label{m:eq4}   k^{e}_{r} = \gamma k_{-1}  - k_3.
\end{eqnarray}

The system  (\ref{m:eq1}), (\ref{m:eq3}), (\ref{m:eq31}) should be supplemented with three conservation laws for concentrations of $R$, $T$ and $A$:
\begin{eqnarray}
\label{m:eq5}
\  R_0 = R + C_R,\\
\label{m:eq51}
\  A_0 = A + C_A, \\
\label{m:eq52}
\  T_0 = T + C_T + C_A + T_i,
\end{eqnarray}
where $R_0$, $T_0$ and $A_0$ are the initial concentrations.

Eqs (\ref{m:eq1}),  (\ref{m:eq3}), (\ref{m:eq31}),
(\ref{m:eq5}) -- (\ref{m:eq52})  form a framework for our analysis.  This is
a  system of nonlinear ODE (because of  conservation laws
(\ref{m:eq5}) -- (\ref{m:eq52}) and because of effective rates  $k^{e}_{f}, k^{e}_{r}$
are functions of the receptor concentration). It can be easily solved
numerically and also allows some analytical progress (see below).
If parameter $\gamma \ll 1$ (and this is the case in many practical
situations), then this model can be reduced to the ``well-mixed''
kinetic model with constant kinetic rates \cite{AR21}.

It is worth emphasizing that the aim our analytical framework is to
develop a simple, but scientifically rigorous model that may be used
to predict the optimum antibody kinetic properties and concentration
required to achieve a desired protective effect, rather than develop
a detailed, biologically accurate model that captures all the detail
of the toxin internalization process. Therefore, the model does not
take into account the pharmacokinetics of the toxin-antibody complex
\cite{R9} or receptor internalization and recycling \cite{R21}, \cite{R22}. $k_3$ is a lumped constant that describes all the processes that result in the appearance of the free toxin in the intracellular space \cite{R23}.  Wiley and Cunningham \cite{R24} and Shankaran et al. \cite{R25} have also developed mathematical models of this type of process.

We are particularly interested in the behaviour of the model under
conditions most likely to reflect the real biological situation i.e.
toxin concentration much lower than the concentration of receptors ($T_0/R_0 \ll 1$).

\begin{table}[h]
     \caption{\label{table:01} Kinetic constants used in numerical simulations (the binding of ricin to its receptor and the monoclonal antibody 2B11).}
\begin{center} 
\begin{tabular}{lccrc}
& Reaction & Value & \\
\hline
& $k_1$ & $1.3 \cdot 10^{5}$& \\
& $k_{-1}$ & $1.4 \cdot 10^{-2}$ & \\
& $k_2$ & $1.25 \cdot 10^{5}$& \\
& $k_{-2}$ & $5.2 \cdot 10^{-4}$& \\
& $k_{-3}$ & $3.3 \cdot 10^{-5}$& \\
\end{tabular}
\end{center}
\end{table}

Testing of the model was carried out using  \texttt{COPASI}
 (software application for simulation and analysis of biochemical networks and their dynamics \cite{R26}) and the kinetic parameters for the binding of ricin to its
receptor and its internalization \cite{R27}  and competition by the
monoclonal antibody 2B11 \cite{R6}. The kinetic parameters used are shown
in Table 1. The value of $k_3$ used is that determined by Sandvig et al. \cite{R19} to be the rate of irreversible binding of ricin to HeLa cells. For simplicity, the simulation was carried out using all reactions taking place in the same compartment.

To illustrate the model we used toxin and receptor concentrations
based on cell culture studies carried out in our laboratory. These
typically use a cell concentration of $1\cdot 10^{4}$ cells per 100 $\mu l$
experiment and a ricin concentration of 10 $pM$. Assuming $3 \cdot  10^{7}$
receptors/cell \cite{R27},  the receptor concentration is approximately
$5 nM$.

\section{Analytical Results}

\subsection{Cell-Surface Binding}

Initially we derive some analytical results for toxins that act at
the cell surface and are not internalized, i.e. we set $k_3 = 0$ in Eq.
(\ref{m:eq31}). At equilibrium $d/dt = 0$   and from (\ref{m:eq1}), (\ref{m:eq3})  we can write
\begin{eqnarray}
\label{m:eq6}  C_R = R T / K_1 ,~~~  C_A = A T / K_2,
\end{eqnarray}
where $K_1 = k_1/k_{-1}, K_2 = k_2/k_{-2}$ are the association
constants for the toxin binding to the receptor and antibody
respectively. It is worth noting that the parameter $\gamma$
(diffusive correction of the intrinsic reaction rates) disappears from
Eqs.(\ref{m:eq6}), so in this case the analytical results are
identical to ones derived using the ``well-mixed'' approximation \cite{AR21}.

In order to simplify notations we denote by $z$ and $y$ the equilibrium
concentrations of the toxin-receptor and toxin-antibody complexes, i.e.
\begin{eqnarray}
\label{m:eq7}   z  = [C_R]_{eq}, ~~ y = [C_A]_{eq}.
\end{eqnarray}
From  Eqs.(\ref{m:eq6})  and conservation laws (\ref{m:eq5})--(\ref{m:eq52}) the following closed equation for $z$ can be derived:
\begin{eqnarray}
\label{m:eq8}  (R_0 - z)(T_0 - z - y) - K_1 z = 0 ,\\
\label{m:eq81}
 y =  A_0  \frac{\epsilon z}{R_0 - z (1 - \epsilon)} ,
\end{eqnarray}
where $\epsilon = {K_1}/{K_2}$.

Eq.(\ref{m:eq8}) can be written in a more conventional form of a cubic equation:
\begin{eqnarray}
\label{m:eq9}   a_3 z^3 + a_2 z^2 + a_1 z + a_0 = 0,
\end{eqnarray}
where
\begin{eqnarray}
\nonumber
 a_3 = \epsilon - 1 ,
 \\
 \nonumber
 a_2 = (1 - \epsilon) C_0 +
\epsilon A_0 + R_0, r
\\
\nonumber
a_1 = - R_0 (C_0 + A_0 + (1 - \epsilon)
T_0),\\
\nonumber
a_0 =  T_0 R^{2}_0,
\end{eqnarray}
and $C_0 = R_0 + K_1$.

It is well-known that Eq.(\ref{m:eq9}) has a closed-form analytical
solution (Cardano's formula \cite{R28}), which in our case provides a consistent way to
derive exact solutions for the proposed model. Unfortunately these solutions still involve rather cumbersome expressions, which require further simplifications in order to be used in practical situations. Below we present another  approach that explicitly employs the smallness of ratio $T_0/R_0 \ll 1$ and leads to a simple analytical expression for the protective properties of the antibody.

We observe that in the absence of antibody (i.e. $A_0 =0$), Eq.(\ref{m:eq8}) is an
elementary quadratic equation that has two roots. If we impose the
obvious constraint $z \rightarrow 0$  as $T_0 \rightarrow 0$ then
there is only one solution, which we designate as $z_0$:
\begin{eqnarray}
\label{m:eq10}z_0 = \frac{C_0}{2} \left( 1- \left( 1 - \frac{4 R_0 T_0}{ C^{2}_0} \right)^{1/2}\right ).
\end{eqnarray}
Under the condition $T_0/R_0 \ll 1$, this can be simplified to
\begin{eqnarray}
\label{m:eq11}
   z_0 \approx \frac{R_0 T_0}{C_0}, ~~ C_0 = R_0 + K_1.
\end{eqnarray}

Let us now evaluate the effect of adding an antibody.  From a
mathematical point of view this effect (i.e. change of $z$ under condition $A_0 > 0$)
is  captured entirely  by the term $y$ in (\ref{m:eq8}), so our aim is to
provide a reasonable analytical estimation of this term.

From Eq.(\ref{m:eq81}) and based on our initial assumption of low toxin concentration ($T_0/R_0 \ll
1$) we can deduce the following simple estimate $y\approx \epsilon z
A_0/R_0$. This then leads to a modified form of Eq.(\ref{m:eq8})
\begin{eqnarray}
\label{m:eq13}  (R_0 - z)(T_0 - z) - K_* z = 0,
\end{eqnarray}
where
\begin{eqnarray}
\label{m:eq14}   K_* = K_1 + \epsilon A_0.
\end{eqnarray}

We can see that this is the same form as the equation for $z$ when
$A_0 =0$, but now with $K_1$ replaced with $K_*$. This also implies
that the analytical solution (\ref{m:eq11}) is still valid, but only
with the substitution $K_1 = K_*$.

In order to characterize the effect of an antibody on the binding of
a toxin to its receptor, we introduce the non-dimensional parameter
$\Psi$, the relative reduction in $C_R$ due to the introduction of
an antibody
\begin{eqnarray}
\label{m:eq15} \Psi   \equiv  \frac{z(A_0 > 0)}{z(A_0 = 0)}.
\end{eqnarray}
The analytical results presented above enable us easily to derive a simple
formula for the antibody efficiency parameter $\Psi$. By using
(\ref{m:eq7}),  \ref{m:eq11}, (\ref{m:eq14}), (\ref{m:eq15}) we can readily deduce
\begin{eqnarray}
\label{m:eq16} \Psi  = \frac{1}{1 +  \epsilon \lambda },~~ \epsilon
=K_1/K_2,~~\lambda = A_0/C_0.
\end{eqnarray}
This expression is the main result of the current paper and will be validated with numerical simulations.

To conclude this section let us briefly discuss some additional constraints for the parameters of our model in order for the expression (\ref{m:eq16}) to be valid. As  mentioned above the condition of low toxin concentration is always assumed in our study. Another simple condition can be derived from the constraint $C_R + C_A \leq T_0$ and by using Eq.(\ref{m:eq11}):
\begin{eqnarray}
\label{m:eq17} \frac{R_0}{C_0} (1 + \epsilon\frac{A_0}{C_0}) \approx \epsilon \frac{R_0 A_0}{C^{2}_0} \leq 1,
\end{eqnarray}
since $R_0/C_0 \leq 1$. This condition could be always checked retrospectively and always hold in our numerical simulations.


\subsection{Toxin Internalization}

For toxins that are internalized, the effect of antibodies that prevent receptor binding is to reduce the effective rate of internalization.  To examine and evaluate this effect we need to analyze the full system (\ref{m:eq1}), (\ref{m:eq3}), (\ref{m:eq31}).

In order to characterize the effect of antibody concentration on the rate of toxin internalization we introduce a new parameter:
\begin{eqnarray}
\label{TI:1}
        G  = \frac{T_i(A_0 > 0)}{T_i(A_0 = 0)},
\end{eqnarray}
which is a function of time (i.e. $G \equiv G(t)$).

Our aim is to deduce  function $G$  based on the kinetic model (\ref{m:eq1}), (\ref{m:eq3}), (\ref{m:eq31}). It is evident that $G \leq 1$ for $t > 0$ and $G \rightarrow 1$ as $t \rightarrow  \infty $ (since in that case all toxin will be internalized).

For the toxins of interest, while the receptor binding is rapid (time sale $\sim 1/(k_1 C_0)$) \cite{A1}, \cite{A2}, the subsequent internalization is much slower (time scale  $\sim  1/k_3 \gg  1/(k_1 C_0)$). This coupling of slow and fast processes in our system allows us to develop a simplified model of toxin internalization using the the well-known framework of Quasi-Steady-State Approximation (QSSA), see \cite{A1}-- \cite{A6} and refs therein.

 When applied to our system  QSSA elucidates the toxin internalization as a two-stage process. After the initial rapid binding of the toxin to the receptor we can simply set $d C_R /dt = 0$ in (\ref{m:eq1}). The further slow evolution of $T(t)$ (viz, quasi-steady state) is completely determined by the conservation law (\ref{m:eq52}) and Eq.(\ref{m:eq31}) and  spans a time scale of the order of the internalization time ($\sim  1/k_3$).  In addition, for solving (\ref{m:eq31}) at the initial stage of internalization, we can assume that $T_i \ll T_0$ and write
\begin{eqnarray}
\label{TI:2}
        T_i (t)  = k_3 z_0 t,  ~~ t \ll 1/k_3,
\end{eqnarray}
where $z_0$ is given by expressions (\ref{m:eq10}) and (\ref{m:eq11}). The evolution of $T_i(t)$ for the late stage of internalization can be readily derived from (\ref{m:eq31}), (\ref{m:eq5}) -  (\ref{m:eq52}) by assuming $[T_0 - T_i(t)]\ll T_0$:
\begin{eqnarray}
\label{TI:3}
        T_i (t)  = T_0 [1 - \exp (- k_3 t)], ~~t \geq  1/k_3,
\end{eqnarray}
 so $T_i (t)$ exponentially approaches its saturation limit. A simulation of this process is shown in Fig.\ref{F:4} and the slow linear increase of $T_i$ at the initial stage is clearly visible.

Now, consider the case where $A_0 > 0$. According to (\ref{TI:2}) the main effect of the introduction of an antibody is to reduce the value of $z_0$, as described in the previous section. Then, based on  (\ref{TI:1}), (\ref{TI:2})  and  (\ref{m:eq15}) we can conclude that, during the quasi-equilibrium stage, the following approximation holds
\begin{eqnarray}
\label{TI:4}
        G  = \frac{T_i(A_0 > 0)}{T_i(A_0 = 0)} \approx \Psi,
\end{eqnarray}
where $\Psi$ is given by expression (\ref{m:eq16}).

The overall effect of introducing an antibody can be  best describes in terms of the internalization half-time, $\tau_{i}$. Without antibody the later can be estimated from (\ref{TI:3}) and  condition $T_i(\tau_{i}) = T_0/2$. Thus from Eq.(\ref{TI:2}) we yield
\begin{eqnarray}
\label{TI:5}
          \tau_{i}  \approx \frac{T_0}{2 k_3 z_0}  = \frac{C_0}{2 k_3 R_0}.
\end{eqnarray}
For the internalization time with the presence of antibody we can apply reduced value of $z_0$  and  write the following simple formula
\begin{eqnarray}
\label{TI:6}
          \frac{\tau_{i}}{\tau^{0}_{i}} \approx \frac{1}{\Psi},
\end{eqnarray}
where $\tau^{0}_{i}$ is the internalization time in the absence of antibody ($A_0=0$).

Eqs.(\ref{TI:5}) and (\ref{TI:6}) have a clear interpretation. As described in the previous section, the introduction of an antibody results in a decrease, at $t \ll \tau_{i}$, in the equilibrium value of $C_R$ (i.e. in $z_0$). This can be related, in accordance with Eq.(\ref{TI:2}) and Eq.(\ref{TI:5}), to a corresponding decrease in the concentration of internalized toxin $T_i$ and a consequent increase in the toxin internalization time (since it takes longer to achieve a give level of $T_i$). Since changes in  $z_0$ can be described comprehensively by means of the parameter $\Psi$, it still remains the only parameter needed to characterize the influence of an antibody on the concentration of internalized toxin (\ref{TI:4}), (\ref{TI:6}).

It is evident that the two main effects described above (reduction of the concentration of internalized toxin at a given time, and increase in the time required for the internalized toxin to reach a given concentration) are not independent of each other. The linear relationships (\ref{TI:4}), (\ref{TI:6}) allow us to establish a general identity that relates these two effects for any time $t$.

Let us assume that for $A_0=0$,  $\tau_0$ in is the time taken for the internalized toxin to reach a concentration $T^{0}_i$  (i.e   $\tau^{0} = T^{0}_i/(k_3 z_0)$, see (\ref{TI:2})). The effect of introducing an antibody is to reduce the internalized toxin concentration to a value $T_i \le T^{0}_i $ . Then from (\ref{TI:4}), (\ref{TI:6}) we can derive the following identity:
\begin{eqnarray}
\label{TI:7}
          T_i \tau_{i} = T^{0}_i \tau^{0}_{i},
\end{eqnarray}
where $\tau_{i}$  is the time required for the internalized toxin to reach  $T^{0}_i$ when $A_0 > 0$.
The identity (\ref{TI:7}) has no explicit dependency on antibody kinetic parameters or concentration and provides an easy way to calculate any of the parameters ($T_i, T^{0}_i, \tau_{i}, \tau^{0}_{i}$) if the other three are known.

\section{Numerical Results and Discussion}

We have derived an analytical expression for the parameter $\Psi$, the relative ability of an antibody to reduce the binding of a toxin to its receptor (\ref{m:eq16}). Our derivation is based on the following assumptions:

1. Toxin concentration is much lower than the receptor concentration

2. For toxins that are internalized, the internalization rate is much slower than establishment of the receptor-toxin binding equilibrium.

Applying these assumptions we found that parameter $\Psi$ is independent of the toxin concentration (see (\ref{m:eq16})), i.e. it is determined by the ratio of antibody to receptor concentration and not by the ratio of antibody to toxin concentration as is commonly used. For the low toxin/receptor ratios likely to occur in biological situations, the condition (\ref{m:eq17}) can be met by large range of antibody kinetic parameters.  From this point of view Eq.(\ref{m:eq16}) should be valid for most practical applications.

\begin{figure}
    \centering
    \includegraphics[height= 0.6\textwidth,width= 0.8\textwidth]{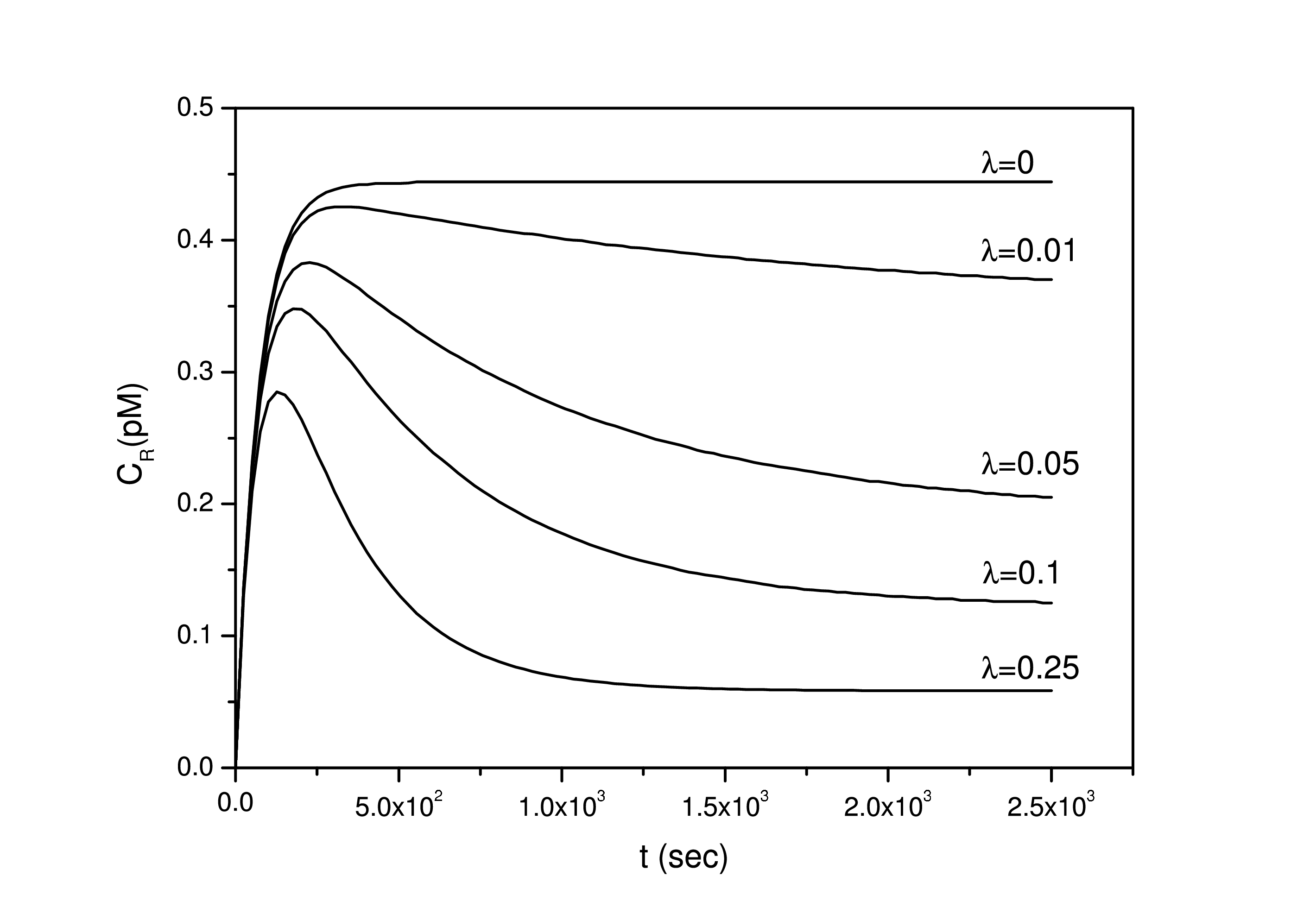}
\caption {Simulated effect of antibody concentration on formation of toxin-receptor complexes $C_R$. Parameter $\lambda = A_0/C_0, C_0 = R_0 + K_1$. The binding curves were created using the simulation package  \texttt{COPASI}  and the kinetic constants in Table 1. $R_0 = 5nM$; $T_0 =10pM$, $C_0=1.15\cdot 10^{-7}$.}
\label{F:1}
\end{figure}

\begin{figure}
    \centering
    \includegraphics[height= 0.5\textwidth,width= 0.7\textwidth]{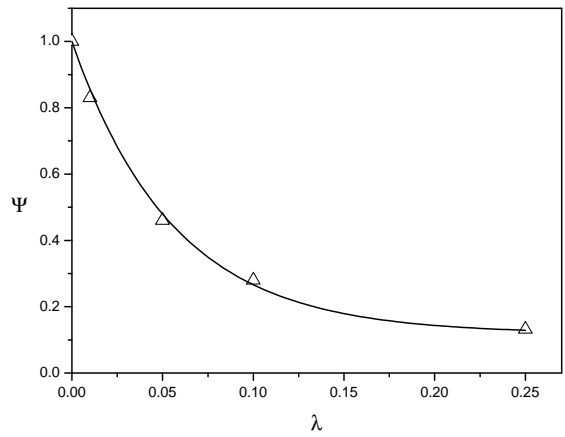}
\caption{Effect of antibody concentration on protection factor. Parameter $\Psi$ (\ref{m:eq15})  was determined  from Eq.(\ref{m:eq16}) (solid lines) and by using simulated values of $C_R$ from Fig.\ref{F:1} at 2500 sec ($\bigtriangleup$), $\epsilon=25.9$.}
\label{F:2}
\end{figure}

\begin{figure}
    \centering
    \includegraphics[height= 0.5\textwidth,width= 0.7\textwidth]{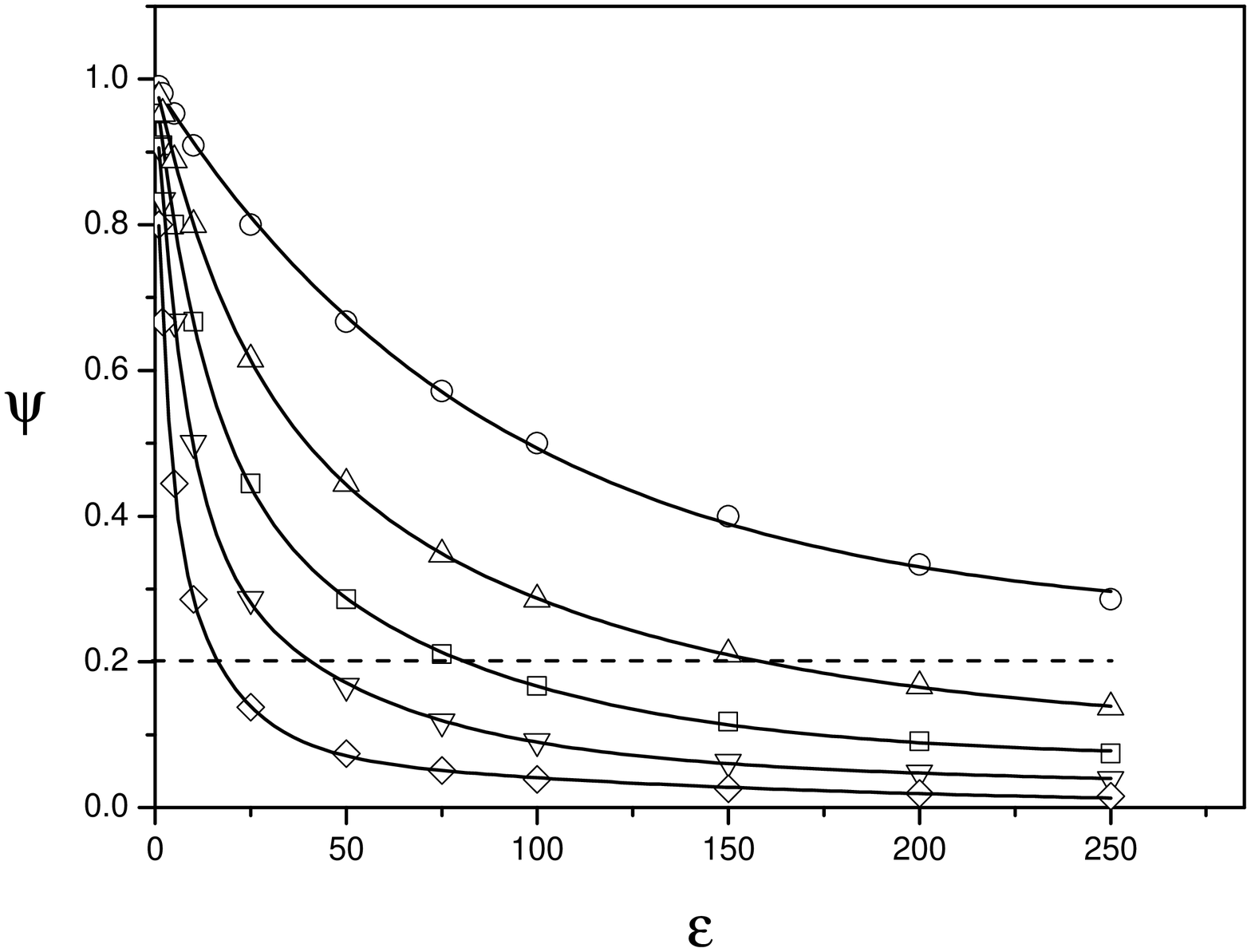}
\caption{Protection factor $\Psi$ (\ref{m:eq15}) as a function of parameter $\epsilon = K_1/K_2$ and $\lambda = A_0/C_0$ (Eq.(\ref{m:eq16})): $\lambda =0.01$ ($\circ$); 0.025 ($\bigtriangleup$); 0.05 ($\square$); 0.1 ($\bigtriangledown$); 0.25 ($\lozenge$). The range of values for $\lambda$ and $\epsilon$  below dashed line corresponds to $80\%$ protection.}
\label{F:3}
\end{figure}

The implications of our analytical findings are illustrated by simulation of the complete kinetic model (Eqs. (\ref{m:eq1}),(\ref{m:eq3}), (\ref{m:eq31}), (\ref{m:eq5})-- (\ref{m:eq52})) using the kinetic constants for ricin and the anti-ricin antibody 2B11 (Table 1).  Fig.\ref{F:1}   is a simulation of the effect of the presence of an antibody on the binding of the toxin to its receptor (formation of $C_R$).  The antibody concentration is expressed as the dimensionless parameter $\lambda = A_0/C_0$.  In this case, since  $R_0$ and $T_0 \ll K_1$, the parameter $C_0 = R_0 + K_1$ is dominated by $K_1$ ($1.08 \cdot 10^{-7}$).

\begin{figure}
    \centering
    \includegraphics[height=0.5\textwidth,width= 0.7\textwidth]{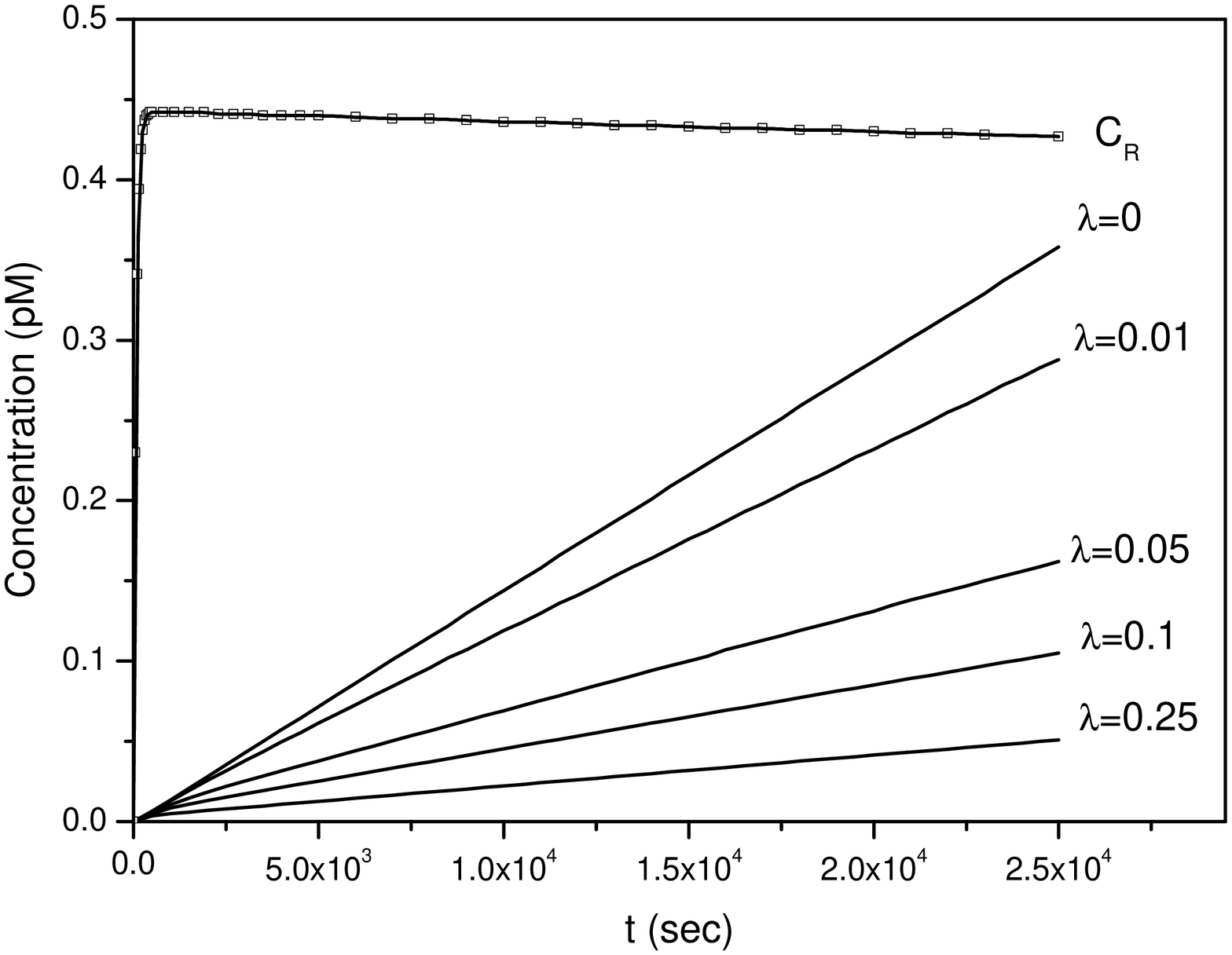}
\caption{Different time scales for formation of receptor-toxin complex  $C_R$ ($\square$) and associated toxin internalisation $T_i$ (solid lines). Results of \texttt{COPASI}  simulation with  kinetic constants from Table 1.  $\lambda = A_0/C_0$, $R_0 = 5 nM$; $T_0=10 pM$, $C_0=1.15 \cdot 10^{-7}$, $\epsilon =25.9$.}
\label{F:4}
\end{figure}

Fig.\ref{F:2} shows the effect of increasing antibody concentration on $\Psi$ . There is a good agreement between the values of $\Psi$   determined from (\ref{m:eq16}) and from (\ref{m:eq15}) using the equilibrium values of $C_R$ determined from simulation of the complete kinetic model (Fig. \ref{F:2}). For instance, the results predict that, for this toxin and antibody combination, the additional protection provided by increasing the antibody concentration diminishes rapidly when $\lambda$ exceeds 0.1.

\begin{figure}
    \centering
    \includegraphics[height= 0.5\textwidth,width= 0.7\textwidth]{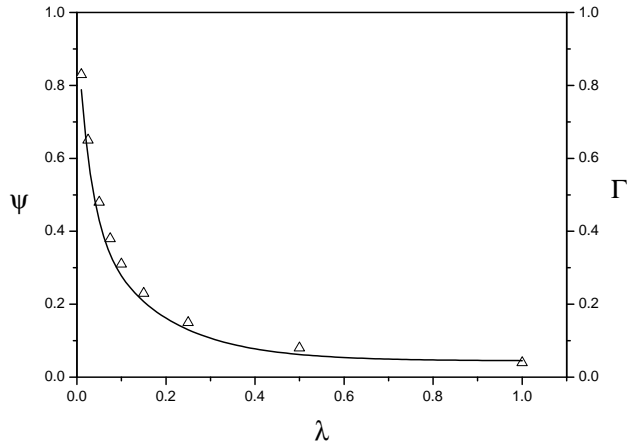}
\caption{Comparison of parameters $\Psi$  and $\Gamma$.  $\Gamma$  ($\bigtriangleup$) was determined using values of $T_i$ and $T^{0}_i$  at $t=10^4$ sec from toxin internalization time courses simulated using  \texttt{COPASI}  and the kinetic constants in Table 1. Parameter $\Psi$  (solid line) was determined from  Eq.(\ref{m:eq16}). $R_0 = 5 nM$; $T_0=10 pM$, $C_0=1.15 \cdot 10^{-7}$, $\epsilon =25.9$.}
\label{F:5}
\end{figure}

Fig.\ref{F:3} shows the relationship (\ref{m:eq16}) between $\Psi$, antibody concentration and the toxin/antibody and the ratio of toxin/receptor dissociation constants ($\epsilon$). This plot is valid for all combinations of toxin, receptor and antibody consistent with the assumptions used to derive (\ref{m:eq16}), principally $T_0\ll R_0$. The antibody kinetic parameters and concentration required to provide a specified degree of protection may be determined from this plot.  For example, any combination of $\epsilon$ and $\lambda$ falling below the dashed line will reduce either $C_R$ or $T_i$ by 80 \%.
\begin{figure}
    \centering
    \includegraphics[height= 0.5\textwidth,width= 0.7\textwidth]{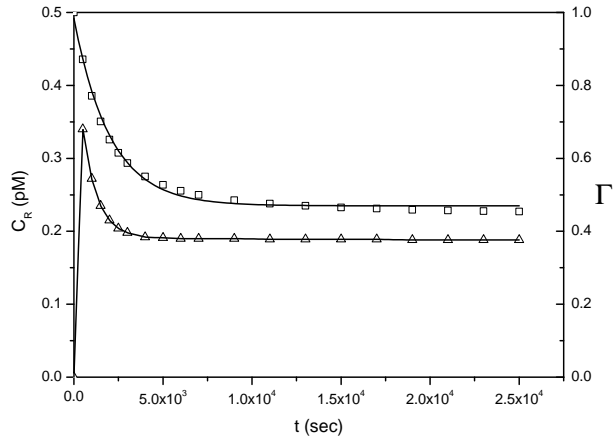}
\caption{Establishment of the quasi-equilibrium state in the presence of antibody. $C_R$ formation ($\triangle$) was
simulated using  \texttt{COPASI} and the kinetic constants in Table 1.  $\Gamma$ ($\square$) was determined using Eq.(\ref{TI:4}) and  values $T_i$ and $T^{0}_i$  at $t=10^4 sec$ using simulated  toxin internalization time courses. $R_0 = 5 nM$; $T_0=10 pM$, $C_0=1.15 \cdot 10^{-7}$, $\lambda=0.05$.}
\label{F:6}
\end{figure}

This, in turn, enables important judgements to be made about antibody selection. For example, if an antibody concentration of $0.25C_0$ ($\lambda =0.25$) is achievable, then an antibody with an $\epsilon$ value of 50 will provide good protection (93\% reduction in $C_R$ or $T_i$). If an antibody concentration of  only $0.05 C_0$ ($\lambda=0.05$) is achievable, then an $\epsilon$ value of  250 is required to achieve the same level of protection. The structure of (\ref{m:eq16}) is such that, a given increase in protection ($\Psi$  or $\Gamma $) may be achieved by either an x-fold increase in $\epsilon$ or an x-fold increase in $\lambda$.

The effect of antibody on toxin internalization is simulated in Fig.\ref{F:4}. Rapid equilibration of receptor and toxin is followed by slow accumulation of toxin within the cell. Equation (\ref{TI:4}) predicts that $\Psi$  is the only parameter needed to characterize the influence of an antibody on toxin internalization. Fig.\ref{F:5} compares  $\Gamma$  calculated using  (\ref{TI:4}), (\ref{m:eq16}) with $\Gamma$  determined using values of $T_i$  and $T^{0}_i$  at $t=10^4 sec$ from this simulated data and shows good agreement between the two values under the condition $T_0 \ll R_0$, although the value of $\Gamma$  is slightly greater than $\Psi$. The plot predicts the degree of protection provided by a given concentration of antibody and enables assessment of the value of increasing antibody concentration beyond a certain value. For example, to enhance the reduction of $T_i$ from 90\% to 95\% requires doubling of $A_0$.

The expression for $\Psi$, (\ref{m:eq16}), assumes a quasi-equilibrium state in the system. In practice, this state may take significant time to achieve. Fig. \ref{F:6} shows a simulation of the time taken by the ricin/receptor/2B11 system to reach the quasi-equilibrium state for $\lambda =0.05$. The value of $\Gamma$   determined from the  toxin internalization profiles (Fig.\ref{F:6}) parallels this process, i.e. experimental validation of $\Gamma$  must allow sufficient time to elapse for the quasi-equilibrium state to be established.

\begin{figure}
    \centering
    \includegraphics[height= 0.5\textwidth,width= 0.7\textwidth]{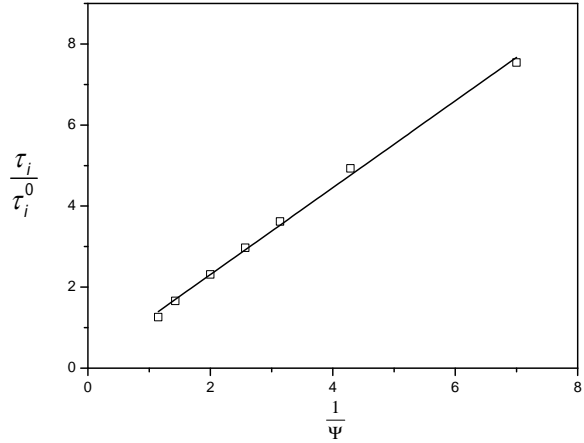}
\caption{Relationship between toxin internalization time $\tau_i$  and protection factor $\Psi$ (\ref{m:eq15}). Solid line is formula (\ref{TI:6}) and ($\square$) is simulation with \texttt{COPASI}. $\tau_i$  was determined as the time to internalize $5 \cdot10^{-14} M$ of ricin. All other parameters are the same as in Fig.\ref{F:6}.}
\label{F:7}
\end{figure}

The relationship between the internalization time $\tau_i$ and $\Psi$  described in Eq.(\ref{TI:6}) is shown in Fig.\ref{F:7}. $\Psi$  was determined from simulated toxin internalization time courses (Fig.\ref{F:4})  as the time to internalize $5 \cdot 10^{-14} M$ ricin. The slope of the fitted line is 1.07, close to the predicted value of 1.0.

In summary, the protection provided by an antibody against toxins that act either at the cell surface or after binding to the cell surface followed by internalization may be predicted from a simple kinetic model. Protection parameter $\Psi$  is a simple function of antibody, receptor and toxin concentrations and the kinetic parameters governing the binding of the toxin to the receptor and antibody:
\begin{eqnarray}
\label{R:5}
          \Psi = \frac{1}{ 1 + (K_1/K_2)(A_0/C_0)} .
\end{eqnarray}

The calculated value of $\Psi$  matches closely the degree of protection determined from numerical simulation of the binding and internalization reactions and provides a convenient method for predicting the optimum antibody parameters (concentration and dissociation constant) needed to provide effective treatment or prophylaxis for toxins.

\section{Acknowledgements}

The authors acknowledge helpful discussions with Dr Chris Woodruff and Dr Ralph Leslie.








\end{document}